\documentclass{aa}

\usepackage{graphicx}
\usepackage{txfonts}
\usepackage{graphicx}	
\usepackage{amssymb}	
\usepackage[nointegrals]{wasysym} 
\usepackage{hyperref}					
\hypersetup{colorlinks,
	linkcolor=blue,%
	citecolor=blue}

\usepackage{multirow}
\usepackage{diagbox} 
\usepackage{tabularx}
\usepackage{subfigure}

\begin{document}

\title{Centaurs potentially in retrograde co-orbit resonance with Saturn}


   \author{Miao Li
          \inst{1}
          \and
          Yukun Huang\inst{1}
          \and
          Shengping Gong\inst{1}\fnmsep
          \thanks{e-mail:gongsp@tsinghua.edu.cn}
          }

   \institute{School of Aerospace Engineering, Tsinghua Universuty, Beijing, 100084, China\\
             \email{gongsp@tsinghua.edu.cn}
             }

   \date{Received XXX; accepted YYY}


\abstract
{}
{2015 BZ509 is the first asteroid confirmed to be in retrograde co-orbit resonance (or 1/-1 resonance) with the giant planets in the solar system. {While Saturn is the only giant planet whose Trojans are not discovered until now, we identify some small bodies among Centaurs and Damocloids that are potentially in 1/-1 resonance with Saturn in the present study}.}
{We integrate numerically the motion of the 1000 clones (include the nominal orbit)
of each Centaur whose orbit has a semi-major axis between 9.3 au and 9.8 au and an inclination $i > 90\degr$.
To confirm and evaluate the 1/-1 resonant configurations mentioned above, we introduce an useful one degree integrable approximation for planar 1/-1 resonance.}
{We identify four candidates potentially in 1/-1 resonance with Saturn.
The capture in this particular resonant state during the 40000 yr integration timespan is very common for 2006 RJ2 (906/1000 clones), 2006 BZ8 (878/1000 clones), and 2017 SV13 (998/1000 clones), and it is less likely for 2012 YE8 (426/1000 clones).
According to our statistical results, 2006 RJ2 is the best candidate to be currently in a 1/-1 mean motion resonance with Saturn, and 2017 SV13 is another important potential candidate. Moreover, 2012 YE8 and 2006 BZ8 are also Centaurs of interest but their current and long-term 1/-1 resonant state with Saturn is less likely.
The proportions of the clones captured in the relative long-term stable co-orbit resonance (over 10000 yr) are also given.
The motions of the 2006 RJ2, 2015 BZ509, and 2006 BZ8 in the solar system are just around the ideal equilibrium points of the 1/-1 resonance given by the planar semi-analytical model.}
{Small bodies in retrograde co-orbit resonance with giant planets may be more common than previously expected. Identification of these potential mysterious minor bodies encourages the search for such objects on a larger scale in our solar system.
The findings of this paper are also useful for understanding the origin and dynamical evolution of the Centaurs and Damocloids on retrograde orbits.}


   \maketitle
%

\section{Introduction}

All of the major planets and the significant portions of the asteroids have orbits around Sun with inclination $i<90\degr$ when viewed from above the north ecliptic pole of our solar system \citep{Wiegert2017}. But small bodies of the solar system on retrograde orbits do exist, such as irregular satellites which have eccentric, highly
inclined or even retrograde orbits around their giant planets \citep{Jewitt2003}.
Although the origins of these mysterious retrograde objects are not clear until now, they are generally believed to have been captured from heliocentric orbit during the final phase of planetary accretion \citep{Pollack1979,Heppenheimer1977}.
Halley-type comets (HTCs) provide us some good candidates to study the dynamics of retrograde objects, while their origin has long been a matter of debate until now
\citep{doi:10.1046/j.1365-8711.1998.01628.x, 1999esra.conf..327F, WIEGERT199984, doi:10.1046/j.1365-8711.2002.05460.x, 2014Icar..231...99F, doi:10.1093/mnras/234.2.389, 1538-3881-121-4-2253,LEVISON2006619, doi:10.1093/mnras/stw1532}.
{Minor bodies} in retrograde orbits are rare in our solar system. Among the 753782 {small bodies} found so far, only 93 are in retrograde motion\footnote{Minor Planet Center; \url{https://www.minorplanetcenter.net/iau/MPCORB.html}, retrieved 18 January 2018.}.
In recent years, the research about the heliocentric retrograde orbital motion gradually multiplies but lots of details remain uncertain until now.
\citet{doi:10.1093/mnrasl/slt106} identified some asteroids trapped into retrograde resonance for thousands of years with Jupiter and Saturn.
\citet{2041-8205-749-2-L39} suggested that the retrograde orbits inside the Jupiter's heliocentric path may be produced through the gravitational interaction of Jupiter and Saturn.
Surprisingly, after the discovery of the first retrograde extrasolar planet, which is known as HAT-P-7b \citep{Narita2009, Winn2009},
\citet{Triaud2010} identified some other exoplanets on retrograde orbits.
Interestingly, \citet{2041-8205-827-2-L24} identified a new retrograde trans-Neptunian object (TNO), nicknamed Niku, which is confirmed in 7/9 mean motion resonance with Neptune by \citet{doi:10.1093/mnrasl/slx125}.
These findings further stimulate the researchers' interest in retrograde motion.
Understanding the dynamics of the retrograde minor bodies is significant for the study of
the mechanisms and origins of the major and minor planets on retrograde orbits.

More than 7000 Trojans have been discovered until now, which are locked in stable prograde co-orbit motion with Jupiter.
Some Trojans of other planets including Earth, Mars, Uranus and Neptune\footnote{Until now, 1 Earth Trojan, 9 Mars Trojans, 1 Uranus Trojan and 17 Neptune Trojans have been discovered. Data taken
	from website of JPL Small-Body Database Search Engine, retrieved 5 May 2018.} have also been observed in recent years such as the first Trojan of Earth, 2010 TK7 \citep{Connors2011},
but no Saturn Trojans have never been identified \citep{doi:10.1093/mnras/stt1974}.
Interestingly, a retrograde co-orbit motion was not found until recently.
In recent years, some theoretical work on these mysterious orbits has verified that the co-orbit retrograde motion can be stable.
\citet{Morais2013} developed a semi-analytical model for retrograde co-orbit motion through the averaged unexpanded disturbing function. They have confirmed later that
there is a great potential for the retrograde orbits trapped in co-orbit resonance \citep{Namouni2015}.
The stability and properties of the spacial coorbital motion have been studied by \citet{Morais2016} from numerical simulations. The authors confirmed that a stable co-orbit motion can exist regardless of the orbital inclination, which, of course, includes the retrograde case.

The recent discovery of the first asteroid trapped in retrograde co-orbit resonance with Jupiter, 2015 BZ509, has renewed interest in the searching for such mysterious objects \citep{Wiegert2017}. 2015 BZ509 is the first asteroid in retrograde co-orbit resonance with a planet of our solar system, and it has long-term stability.
\citet{Wiegert2017} suggested that this asteroid can keep this stable resonant state about a million years.
The retrograde co-orbit resonance with Jupiter protects it from disruptive close encounters with the other planets.
\citet{Namouni2017} have found that 2015 BZ509 lies almost exactly at the capture efficiency peak by numerical simulations.
Indeed, \citet{doi:10.1093/mnrasl/slt106} have identified that 2006 BZ8 could enter the co-orbit retrograde resonance with Saturn in the future.
From numerical simulations, \citet{Namouni2015} showed that asteroids on retrograde motion are more easily captured in mean motion resonances than those on prograde motion.
In this paper, we identify the dynamics of the {Centaurs} which have co-orbit retrograde orbits around Saturn by numerical simulations,
trying to find some new {small bodies} in co-orbit retrograde resonance with the giant planets of our solar system.

The article is structured as follows: in section 2, we introduce the main properties of the retrograde mean motion resonance{, and the co-orbit retrograde resonance is discussed in detail. Section 3 gives a full description of the numerical model used in our study.} In section 4, {we} show the general statistical results of our numerical simulations and analyze the complicated resonant states we observed. In section 5, we analyze the phase-space portrait of the co-orbit retrograde resonance
	and present a comparison between the results obtained from the semi-analytical model and the numerical simulations. Our conclusions and some discussion are given in the last section.

	\section{Retrograde co-orbit resonance}

	A retrograde mean motion resonance between {a small body} and a planet (denoted by superscript symbol {\arcmin})
	occurs when $pn-qn\arcmin\approx0$, where $p$ and $q$ are both positive integers, and $n$ and $n\arcmin$ are
	the mean motion frequencies of the {small body} and the planet, respectively. We call such resonance
	a retrograde $p/q$ resonance or $p/-q$ resonance.

	To determine whether these {minor bodies} are in retrograde co-orbit resonances with Saturn,
	we have to first get the correct expression of the particular resonant angles defined by the Fourier expansion of the disturbing function \citep{Murray:396121}.
	For retrograde mean motion resonance, \citet{Morais2013} presented a transformation of the relevant resonant term for the corresponding retrograde resonance to keep with the conventional definition of osculating orbital elements in prograde motion.
	The expression of the longitude of the pericenter becomes $\varpi = \omega - \Omega$, while the mean longitude of the asteroid becomes $\lambda = M + \omega - \Omega$, where $\omega$, $\Omega$ and $M$ are the argument of pericenter, the longitude of the ascending node and the mean anomaly, respectively.

	\citet{Morais2013, doi:10.1093/mnrasl/slt106, Morais2016} reported that the resonant angles for the $p/-q$ resonance have the following forms:
	\begin{equation}
		\phi=q\lambda-p\lambda\arcmin-(p+q-2k)\varpi+2k\Omega,
		\label{eq:1}
	\end{equation}
	where $p$, $q$, and $k$ are positive integers. $p+q\ge2k$ must be satisfied, and the lowest order terms are proportional to $\cos^{2k}(i/2)e^{p+q-2k}$, where $i$ and $e$ are respectively the inclination and the eccentricity of the small body.
	The resonant term corresponding to $k=0$ will be the dominant one for the nearly coplanar retrograde motion because of $\cos(i/2)\approx0$.
	In particular, for retrograde co-orbit resonance, the lowest order resonant terms are proportional to
	\begin{equation}
		\begin{split}
			&e^2\cos(\phi), k=0,\\
			&\cos^2(i/2)\cos(\phi_z), k=1,\\
		\end{split}
		\label{eq:2}
	\end{equation}
	where $\phi=\lambda-\lambda\arcmin-2\varpi$ and $\phi_z=\lambda-\lambda\arcmin+2\Omega$ are two resonant angles of the 1/-1 resonance.
	When we study the Centaurs in 1/-1 resonance with Saturn, we also consider the possibility of the Centaurs in 2/-5 resonance with Jupiter as
	the positions of Jupiter and Saturn are near the 5/2 resonance.

	\section{Descriptions about the numerical model}
	To find all of the {small bodies} in the 1/-1 resonance with Saturn, we use the JPL Small-Body Database Search Engine\footnote{JPL Small-Body Database Search Engine; \url{https://ssd.jpl.nasa.gov/sbdb_query.cgi}}
	to search for {small bodies} with retrograde orbits (inclination $i > 90\degr$) and semi-major axes ranging from 9.3 au to 9.8 au (semi-major axis of Saturn $a_{\rm Saturn}$ = 9.54 au).
	There are four Centaurs meet our limitations, which are 2006 RJ2, 2006 BZ8, 2012 YE8 and 2017 SV13.
	{We studied Centaurs} on retrograde orbits,
	which are usually believed strongly chaotic because of their giant-planet crossing orbit.
	The discovered Centaurs mainly have two kinds of dynamical evolution \citep{BaileyBrenaeL2009, Volk2013}: random walk and resonance sticking.
	As these selected {small bodies} among Damocloids (Jupiter Tisserand's parameter $T<2$), \citet{1538-3881-129-1-530} suggested that the most obvious origin
	would be Oort Cloud or Halley-type comet.

	We numerically integrate the motions of these Centaurs in the solar system using the MERCURY \citep{Chambers1999} package with an accuracy parameter $10^{-12}$.
	The Burlisch-Stoer integrator is employed while considering all the gravity of the eight planets throughout the integration.
	As listed in Table~\ref{tab:1} and Table~\ref{tab:2}, the uncertainty parameters $U$\footnote{Uncertainty parameter is defined by Minor Planet Center, on a scale 0-9, 9 being very uncertain.} of 2006 RJ2, 2006 BZ8, 2012 YE8, 2017 SV13 according to Minor Planet Center are 5, 0, 5 and 4, respectively.
	To exclude the influence of uncertainties in the orbit determination, we generated 1000 clone orbits (include the nominal one) for the each selected Centaur by standard deviations (1-$\sigma$ uncertainties), which are corresponding to Gaussian
	distribution around the nominal orbit. We integrate the nominal and clone orbits over a 40000 yr period from -10000 {yr} to 30000 yr around present-day.

\begin{table*}
	\centering
	\caption{Nominal orbital elements at JED2458000.5 of 2006 RJ2.
	The 1-$\sigma$ uncertainty and the Jupiter Tisserand's parameter are also given. Data taken from website of JPL Small-Body Database Search Engine.
	The uncertainty parameter is given by the MPC.}
	\label{tab:1}
	\begin{tabularx}{0.75\textwidth}{m{8cm}m{7cm}}
		\hline
		Full Name                                        & 2006 RJ2                              \\
		\hline
		Epoch (JED)                                      & 2458000.5                             \\
		Semi-major axis, $a$(au)                         & $ 9.676917955\pm9.4898\times10^{-2}$  \\
		Eccentricity, $e$                                & $ 0.7623217118\pm2.2762\times10^{-3}$ \\
		Inclination, $i$(\degr)                          & $164.5713262\pm5.9369\times10^{-3}$   \\
		Longitude of the ascending node, $\Omega$(\degr) & $191.70837\pm2.9035\times10^{-4}$     \\
		Argument of perihelion, $\omega$(\degr)          & $161.421009\pm4.943\times10^{-2}$     \\
		Mean anomaly, $M$(\degr)                         & $133.2269958\pm1.9675$                \\
		Jupiter Tisserand's parameter, $T_J$             & $-1.1674$                             \\
		Uncertainty parameter, $U$                       & 5                                     \\
		\hline
	\end{tabularx}
\end{table*}

\section{Crucial statistical results of numerical simulations}
 To evaluate {this particular} resonant state of each candidate, we focus on the fractions of clones trapped in {1/-1} resonance, currently in {1/-1} resonance and trapped in {1/-1} resonance over 10000 yr {with Saturn}, respectively.
The crucial statistical results of numerical simulations {include the properties of each candidate in 1/-1 mean motion resonance with Saturn} are given in Table~\ref{tab:3}. The following section shows detailed descriptions of dynamical states of the four Centaurs under study.

\begin{figure}
	\includegraphics[width=\columnwidth]{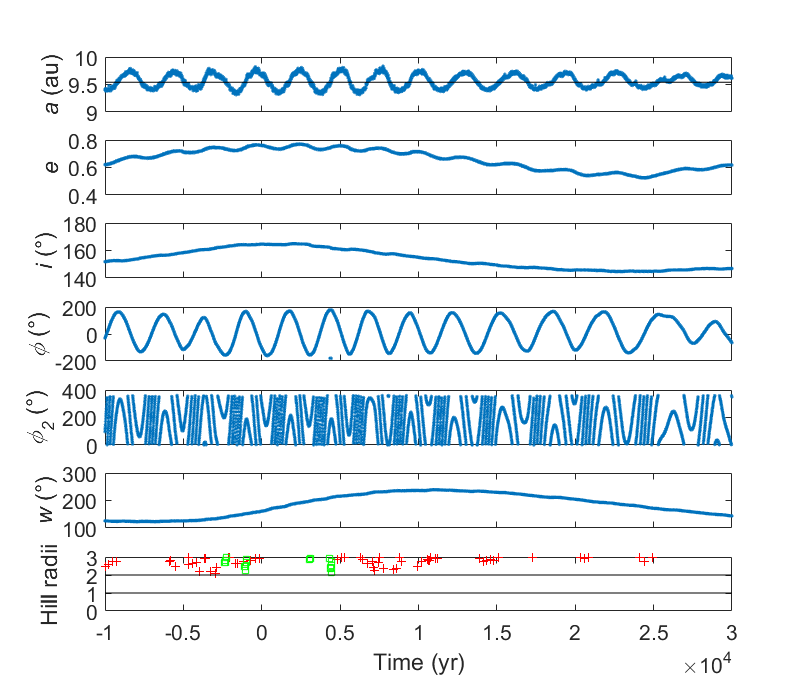}
	\caption{Properties of the nominal orbit of 2006 RJ2. From top to lower panel: semimajor axis $a$ and location of exact 1/-1 resonance with Saturn; eccentricity $e$; inclination $i$;
				angle for nearly coplanar motion involving 1/-1 resonance with Saturn $\phi$; 2/-5 resonant angle with Jupiter $\phi_2$; argument of pericenter $\omega$; the minimum distance in close encounters with Jupiter and Saturn within three Hill's radii of the planet is represented by red plus and green square, respectively.}
	\label{fig1}
\end{figure}

\begin{figure}
	\includegraphics[width=\columnwidth]{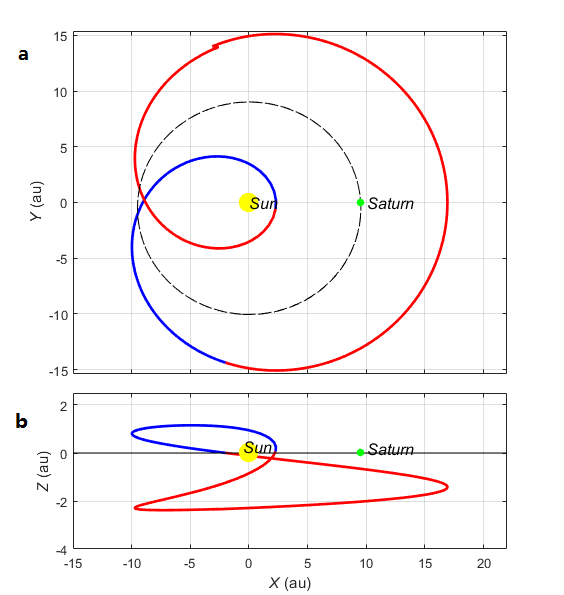}
	\caption{The path of the Centaur 2006 RJ2 in the frame that rotates with Saturn. $X$, $Y$, and $Z$ are the cartesian axes of the Saturn's rotating frame.
				The $X$-axis points from Sun to Saturn and $Z$-axis is vertical to Saturn's orbital plane. Saturn is presented as the green filled circle.
				Panel a): the path of 2006 RJ2 in Saturn's rotating frame viewed from the $Z$-axis direction. The black dotted curve is the Saturn's orbit in the inertial reference frame.
				Panel b): the motion of 2006 RJ2 viewed from the $Y$-axis direction. The blue curve represents the motion of 2006 RJ2 above the Saturn's orbital plane while the red curve means the opposite.}
	\label{fig8}
\end{figure}

2006 RJ2 was discovered on 14 September 2006 and referred to as Centaur. As shown in Table~\ref{tab:1}, the Centaur has an orbit with large inclination and eccentricity.
As shown in Figure~\ref{fig1}, the nominal orbit of 2006 RJ2 is currently in a long-term stable 1/-1 resonance with Saturn. In fact, the fourth panel of Figure~\ref{fig1} shows that
libration of the resonant angle $\phi$ is of large amplitude and is centered on 0\degr, which confirms that the nominal orbit of 2006 RJ2 is in 1/-1 resonance with Saturn.
The oscillation of the resonant angles for most of the clones of 2006 RJ2 also behaves in a similar way.
As shown in Table~\ref{tab:3}, 906 clones are trapped in retrograde co-orbit resonance and 752 clones among them keep this resonant state over 10000 yr.
Above all, a majority of (805/1000) clones are shown to be in the co-orbit resonance at present time.
Therefore, these results seem to suggest that 2006 RJ2 is very likely a retrograde co-orbital Centaur of Saturn,
and that finding is not accidentally due to uncertainties in the orbit determination. {From this, it is probable that 2006 RJ2 is another minor body} apart from 2015 BZ509 in retrograde co-orbit resonance with giant planets.
The last panel of Figure~\ref{fig1} shows that the resonance protects the Centaur from close encounters with the giant planets.
In fact, the minimum distance in a close encounter with the giant planets is always larger than 2 Hill's radii during the whole integration timespan.
The brief capture in the 2/-5 resonance with Jupiter after the {Centaur} exits the 1/-1 resonance is also possible (363/1000 clones).
The path of the {Centaur} 2006 RJ2 in the Saturn's rotating frame is shown in Figure~\ref{fig8}. When viewed from above the plane of Saturn's orbit (Figure~\ref{fig8}a), the motion of 2006 RJ2 has
a typical type of trisectrix curve defined by \citet{Morais2017}.
\citet{Morais2016} have confirmed by the numerical investigations that this trisectrix motion of 2006 RJ2 can be long-term stable if it's influenced only by Saturn.

\begin{figure}
	\includegraphics[width=\columnwidth]{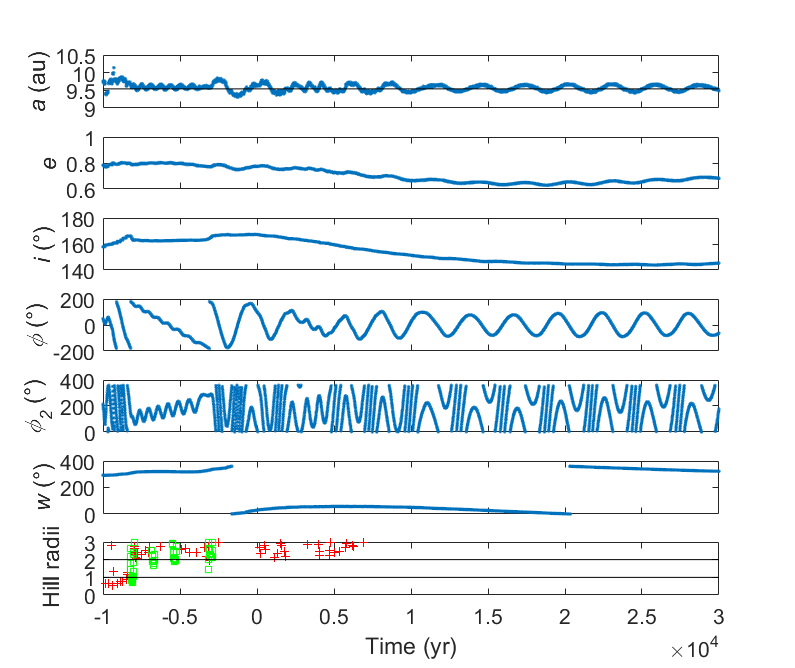}
	\caption{A cloned orbit of 2006 BZ8, which is chosen as one of the best clones captured in 1/-1 resonance. The basic meaning of this figure is the same as Figure~\ref{fig1}.}
	\label{fig3}
\end{figure}

Now, we focus on the other three potential retrograde co-orbital Centaurs of Saturn, 2006 BZ8, 2012 YE8 and 2017 SV13, which have retrograde orbits around Saturn and may be captured into the 1/-1 resonance within the whole integration timespan.
Table~\ref{tab:2} shows their orbital elements and the 1-$\sigma$ uncertainties at JED2458000.5. 2006 BZ8 has been confirmed to be in the 2/-5 resonance with
Jupiter (libration around 180\degr)\citep{doi:10.1093/mnrasl/slt106}.
{Moreover}, as listed in Table~\ref{tab:3}, our numerical statistical results show that this Centaur has low probability of currently in 1/-1 resonant state {with Saturn} (9/1000), which is not mentioned in \citet{doi:10.1093/mnrasl/slt106}.
Moreover, {as shown in Figure~\ref{fig3},} we choose one of the best clones captured in 1/-1 resonant state. {Our statistical results indicate that the capture in 1/-1 resonance with Saturn (libration around 0\degr) within 40000 yr integration timespan is possible (878/1000 clones),}
and 38 clones among them are captured in 1/-1 resonant state over 10000 yr.
\citet{doi:10.1093/mnrasl/slt106} suggested that 2006 BZ8 is in Lidov-Kozai secular resonance, we find that 2006 BZ8 is locked in Lidov-Kozai mechanism (the argument of pericenter $\omega$ librates around 0\degr) at present and will exit at around 120000 yr$^{\ref{foot1}}$.
{The main dynamical properties of the Lidov-Kozai mechanism for retrograde orbits ($i > 90\degr$) outside the mean motion resonances are identical with the symmetrical prograde orbits with inclination $180 - i$.
\citet{1962AJ.....67..591K} studied the secular evolution of a test partical ($a$ = 3 au) in the main asteroid belt while considering the effect of Jupiter only.
\citet{Thomas1996} used the semi-analytical method to study the Lidov-Kozai dynamics around the main belt ($a$ = 3 au) and the Kuiper belt ($a$ = 45 au) by considering the effects of the four giant planets.
\citet{1999A&A...341..928G} introduced an example ($a$ = 7 au) to analyze the Lidov-Kozai dynamics on Centaurs. They pointed out that orbits with moderate value of maximum orbital inclination ($i_{\rm max}$ = 25\degr), the argument of pericenter $\omega$ librates around 0\degr or 180\degr. For even larger $i_{\rm max}$, there will be new libration regions where $\omega = 90\degr$ or 270\degr and large eccentricity.
But there are not systematic studies about the role of Lidov-Kozai mechanism on minor bodies near giant planets until now. We will analyze in detail the effects of Lidov-Kozai dynamics on retrograde Centaurs near giant planets in our future work.}
It's interesting to note that in the restricted three-body model (Sun-Saturn-2006 BZ8), the Centaur 2006 BZ8 is in 1/-1 resonance with Saturn very well over 10000 yr.
It has the similar result in the Sun-Jupiter-2006 BZ8 model where the {Centaur} is in 2/-5 resonance with Jupiter.
We will try to analyze this interesting phenomenon in our future work.

\begin{figure}
	\includegraphics[width=\columnwidth]{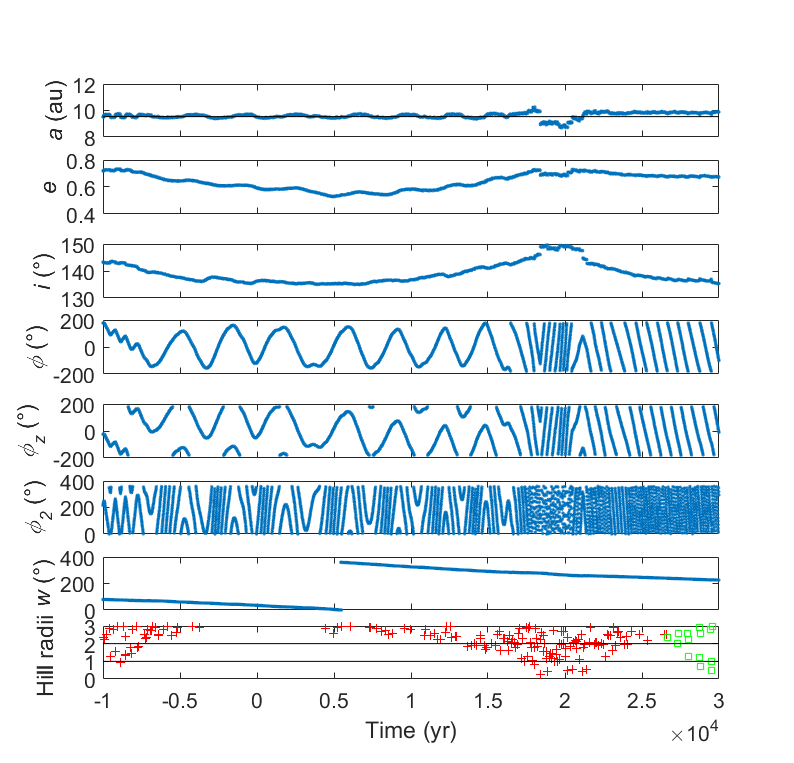}
	\caption{A cloned orbit of 2012 YE8. Since its orbit is far from the planar case, another 1/-1 resonant angle with Saturn $\phi_z$ is included. The other basic meaning of this figure is the same as Figure~\ref{fig1}.}
	\label{fig4}
\end{figure}

2012 YE8 (Figure~\ref{fig4}) is currently near the 2/-5 resonance with Jupiter and the 1/-1 resonance with Saturn at the same time. Close approaches
to Saturn and Jupiter may lead the {Centaur} to enter or exit the resonant states. Since its
orbit is far from the planar case, we consider two possible 1/-1 resonant angles ($\phi$ and $\phi_z$). In our numerical integrations for 2012 YE8, as listed in Table~\ref{tab:3},
140/1000 clones are currently in 1/-1 resonance with Saturn (libration around 0\degr),
426 clones are trapped in this resonant state, and 100 clones among them have relative long-term stability.
Meanwhile, 268/1000 clones are captured in the 2/-5 resonance with Jupiter (libration around 180\degr),
most often occur between $-10000$ yr and $-5000$ yr and from 15000 yr to 25000 yr.
The argument of pericenter $\omega$ always close to 0\degr over the integration timespan, which means a Lidov-Kozai secular resonance\footnote{\label{foot1}Longer integrations have been done to confirm {these} results.}.

\begin{figure}
	\includegraphics[width=\columnwidth]{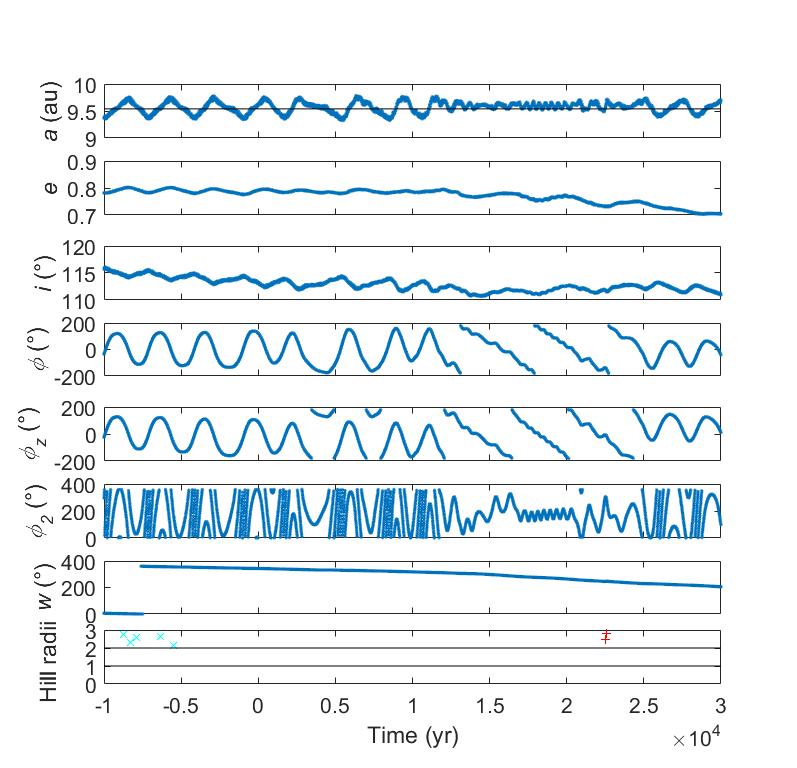}
	\caption{A cloned orbit of 2017 SV13. Since its orbit is far from the planar case, another 1/-1 resonant angle with Saturn $\phi_z$ is included. In the last panel, the minimum distance in close encounters with Jupiter and Uranus within three Hill's radii of the planet is represented by red plus and cyan cross, respectively. The other basic meaning of this figure is the same as Figure~\ref{fig1}.}
	\label{fig5}
\end{figure}

2017 SV13 was just discovered on 17 September 2017 and referred to as Centaur. Different from other three Centaurs, 2017 SV13 has a near-vertical orbit.
As shown in Figure~\ref{fig5}, 2017 SV13 is another suitable candidate in 1/-1 resonance with Saturn.
Almost all of clones (998/1000 clones) are captured in the 1/-1 resonance with Saturn (libration around 0\degr), and the resonant state of 626 clones among them last beyond 10000 yr.
Meanwhile, 434 clones are shown to be in retrograde co-orbit resonance at present, and clones are always expected in 1/-1 resonant state between 5000 yr and 15000 yr timespan.
These results make 2017 SV13 to be another good candidate for Saturn's retrograde co-orbital Centaur.
After leaving the 1/-1 resonance around 15000 yr, 595/1000 clones will be captured in 2/-5 resonance with Jupiter (libration around 180\degr).
The motion is not in Lidov-Kozai secular resonance presently but will be briefly captured in the Lidov-Kozai mechanism (the argument of pericenter $\omega$ librates around  $\pm$90\degr).$^{\ref{foot1}}$

\begin{table*}
	\centering
	\caption{Nominal orbital elements at JED2458000.5 of 2006 BZ8, 2012 YE8, 2017 SV13.
	The 1-$\sigma$ uncertainties and the Jupiter Tisserand's parameter are also given. Data taken from website of JPL Small-Body Database Search Engine.
	The uncertainty parameters are given by the MPC.}
	\label{tab:2}
	\begin{tabularx}{0.75\textwidth}{m{8cm}m{7cm}}
		\hline
		Full Name                                        & 2006 BZ8                              \\
		\hline
		Epoch (JED)                                      & 2458000.5                             \\
		Semi-major axis, $a$(au)                         & $ 9.608985507\pm6.812\times10^{-5}$   \\
		Eccentricity, $e$                                & $ 0.8024503954\pm1.3873\times10^{-6}$ \\
		Inclination, $i$(\degr)                          & $165.318425\pm1.8448\times10^{-5}$    \\
		Longitude of the ascending node, $\Omega$(\degr) & $183.7606180\pm4.5765\times10^{-5}$   \\
		Argument of perihelion, $\omega$(\degr)          & $82.46140331\pm6.4218\times10^{-5}$   \\
		Mean anomaly, $M$(\degr)                         & $134.9664552\pm1.4364\times10^{-3}$   \\
		Jupiter Tisserand's parameter, $T_J$             & $-1.027$                              \\
		Uncertainty parameter, $U$                       & 0                                     \\
		\hline
		Full Name                                        & 2012 YE8                              \\
		\hline
		Epoch (JED)                                      & 2458000.5                             \\
		Semi-major axis, $a$(au)                         & $ 9.38168449\pm7.1464\times10^{-2}$   \\
		Eccentricity, $e$                                & $ 0.5914212664\pm3.037\times10^{-3}$  \\
		Inclination, $i$(\degr)                          & $136.1071898\pm2.2571\times10^{-2}$   \\
		Longitude of the ascending node, $\Omega$(\degr) & $135.3614075\pm2.6817\times10^{-2}$   \\
		Argument of perihelion, $\omega$(\degr)          & $35.05260406\pm3.5294\times10^{-2}$   \\
		Mean anomaly, $M$(\degr)                         & $134.9664552\pm5.6912\times10^{-1}$   \\
		Jupiter Tisserand's parameter, $T_J$             & $-1.006$                              \\
		Uncertainty parameter, $U$                       & 5                                     \\
		\hline
		Full Name                                        & 2017 SV13                             \\
		\hline
		Epoch (JED)                                      & 2458000.5                             \\
		Semi-major axis, $a$(au)                         & $ 9.686282362\pm3.4318\times10^{-2}$  \\
		Eccentricity, $e$                                & $ 0.7928247257\pm7.1768\times10^{-4}$ \\
		Inclination, $i$(\degr)                          & $113.2479445\pm6.9472\times10^{-3}$   \\
		Longitude of the ascending node, $\Omega$(\degr) & $11.70939482\pm7.4949\times10^{-4}$   \\
		Argument of perihelion, $\omega$(\degr)          & $343.2582508\pm1.0814\times10^{-2}$   \\
		Mean anomaly, $M$(\degr)                         & $0.6024677028\pm4.0766\times10^{-3}$  \\
		Jupiter Tisserand's parameter, $T_J$             & $-0.119$                              \\
		Uncertainty parameter, $U$                       & 4                                     \\
		\hline
	\end{tabularx}
\end{table*}

\begin{table*}
	\centering
	\caption{Numerical statistics about the fraction of clones for each of the four Centaurs under study.
				The fraction is the ratio of the total clones in 1/-1 resonance to 1000.
				To evaluate the resonant state of each Centaur, we focus on the fractions of clones trapped in resonance (top row), currently in resonance (middle row), and trapped in resonance over 10000 yr (bottom row), respectively}.
	\label{tab:3}
	\renewcommand\arraystretch{1.25}
	\begin{tabularx}{0.89\textwidth}{m{5cm}m{2.5cm}m{2.5cm}m{2.5cm}m{2.5cm}} 
		\hline
		Centaurs                                 & 2006 RJ2 & 2006 BZ8 & 2012 YE8 & 2017 SV13 \\
		\hline
		trapped in 1/-1 resonance                & 906/1000 & 878/1000 & 426/1000 & 998/1000  \\
		currently in 1/-1 resonance              & 805/1000 & 9/1000   & 140/1000 & 434/1000  \\
		captured in 1/-1 resonance over 10000 yr & 752/1000 & 38/1000  & 100/1000 & 626/1000  \\
		\hline
	\end{tabularx}
\end{table*}

\section{Phase-space portrait of 1/-1 mean motion resonance}
To confirm that the {Centaurs} may be captured in 1/-1 resonance with Saturn, this section is devoted to the construction of the theory of the retrograde mean motion resonance, and the phase-space portrait of 1/-1 mean motion resonance specifically.
{Follow the method introduced by \citet{huang2018dynamic} to analyze the dynamics of 1/-1 resonance}, firstly, we introduce the following retrograde Poincare variables:
\begin{equation}
	\begin{split}
		&\Lambda=L, \ \ \ \ \ \ \ \ \ \ \ \ \ \ \ \ \ \lambda=M+\omega-\Omega,\\
		&P=L-G, \ \ \ \ \ \ \ \ \ \  p=-\omega+\Omega,\\
		&Q=G+H, \ \ \ \ \ \ \ \ \  q=\Omega,\\
	\end{split}
	\label{eq:3}
\end{equation}
where $L=\sqrt{a}$, $G=L\sqrt{1-e^2}$, $H=G$$\cos$$i$. The canonical variables $L,M,G,\omega,H,\Omega$ are usually called the Delaunay variables. These new
		retrograde Poincare variables are different from these given by \citet{Murray:396121}. But it is easy to check that the following contact transformation
		relationship holds:
		\begin{equation}
			\Lambda \lambda+P p+Q q=LM+G\omega+H\Omega,
			\label{eq:4}
		\end{equation}
		So the new retrograde Poincare variables is a set of canonical action-angle variables. The 1/-1 mean motion resonant Hamiltonian for the circular
		restricted three-body model is given by \citet{2002mcma.book.....M}
		\begin{equation}
			\mathcal{H}=\mathcal{H}_0(\Lambda,\Lambda\arcmin)+\epsilon\mathcal{H}_1(\Lambda,P,Q,p,q,\lambda-\lambda\arcmin),
			\label{eq:5}
		\end{equation}
		where
		\begin{equation}
			\mathcal{H}_0=-\tfrac{1}{2\Lambda^2}+n\arcmin\Lambda\arcmin,
			\label{eq:6}
		\end{equation}
		is the main term, and $\mathcal{H}_1$ is the disturbing term. To study the dynamics of Hamiltonian~(\ref{eq:5}), we introduce a set of
		canonical action-angle variables as follows:
		\begin{equation}
			\begin{split}
				&S=P, \ \ \ \ \ \ \ \ \ \ \ \ \ \ \ \ \ \ \ \ \ \ \ \  \ \ \ \ \ \sigma=\tfrac{\phi}{2},\\
				&S_z=Q, \ \ \ \ \ \ \ \ \ \   \ \ \ \ \  \ \ \ \ \ \ \ \ \ \ \ \sigma_z=\tfrac{\phi_z}{2},\\
				&N=-2\Lambda+P+Q, \ \ \ \ \ \ \ \ \  \upsilon=-\tfrac{\lambda-\lambda\arcmin}{2},\\
				&\tilde{\Lambda}\arcmin=\Lambda\arcmin+2\Lambda, \ \ \ \ \ \ \ \ \  \ \ \ \ \ \ \  \tilde{\lambda}\arcmin=\lambda\arcmin,\\
			\end{split}
			\label{eq:7}
		\end{equation}
		where $\sigma$ and $\sigma_z$ are the critical angles of the retrograde mean motion resonance.
		If we restrict our retrograde motion to planar case $i=180\degr$, the second resonant angle $\phi_z$ is neglectable and then $S_z$ or $Q$ is always equal to 0.
		Besides, it is obvious that $N$ is constant of motion for $\mathcal{H}$, {whose expression is now}
		\begin{equation}
			N=-2\Lambda+P+Q=-\sqrt{a}(1+\sqrt{1-e^2}),
			\label{eq:8}
		\end{equation}
		which define a curve in the $a-e$ plane. Thus the Hamiltonian of the planar case only depends on one critical angle $\sigma$. The new integrable Hamiltonian reads:
		\begin{equation}
			\mathcal{H}_{\rm planar}=\mathcal{H}_0(N,S)+\epsilon\mathcal{H}_1(N,S,2\sigma),
			\label{eq:9}
		\end{equation}

\begin{figure}
	\includegraphics[width=\columnwidth]{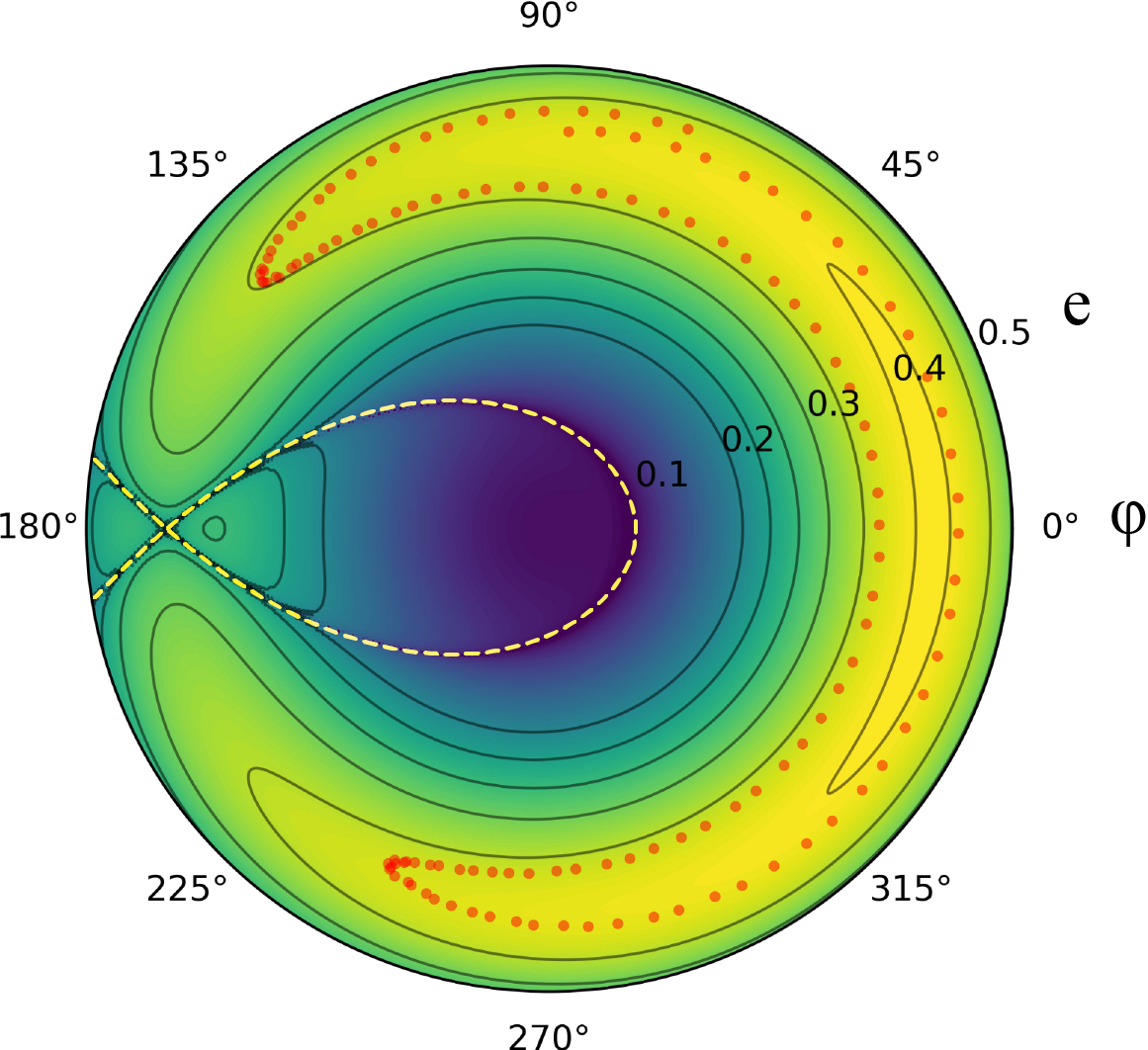}
	\caption{The phase-space portrait of the 1/-1 mean motion resonance with Jupiter associated with the orbital elements of the mysterious asteroid 2015 BZ509, presented in the  $e-\phi$ polar space. The
	variance of color and contours represent level curves of the $\mathcal{H}_{\rm planar}$ (equation~\ref{eq:9}).
	The red scatter diagram present the results of the numerical integration in the N-body numerical model,
	while the yellow dashed line denotes the collision curve with the perturber.}
\label{fig7}
\end{figure}

As for the disturbing term $\epsilon\mathcal{H}_1$, we numerically evaluate its value by averaging over all fast angles \citep{2002mcma.book.....M}. Then we can calculate
the value of equation~(\ref{eq:9}) on $e-\phi$ plane.
We apply this planar 1/-1 resonance semi-analytical model to the first asteroid in 1/-1 resonance with Jupiter, 2015 BZ509\footnote{Orbital elements of 2015 BZ509 taken from JPL Small-Body
	Database Search Engine; \url{https://ssd.jpl.nasa.gov/sbdb_query.cgi}, retrieved on January 2018.}, which has a nearly coplanar orbit.
We can predict its motion in numerical model very well (Figure~\ref{fig7}).

\begin{figure}
	\includegraphics[width=\columnwidth]{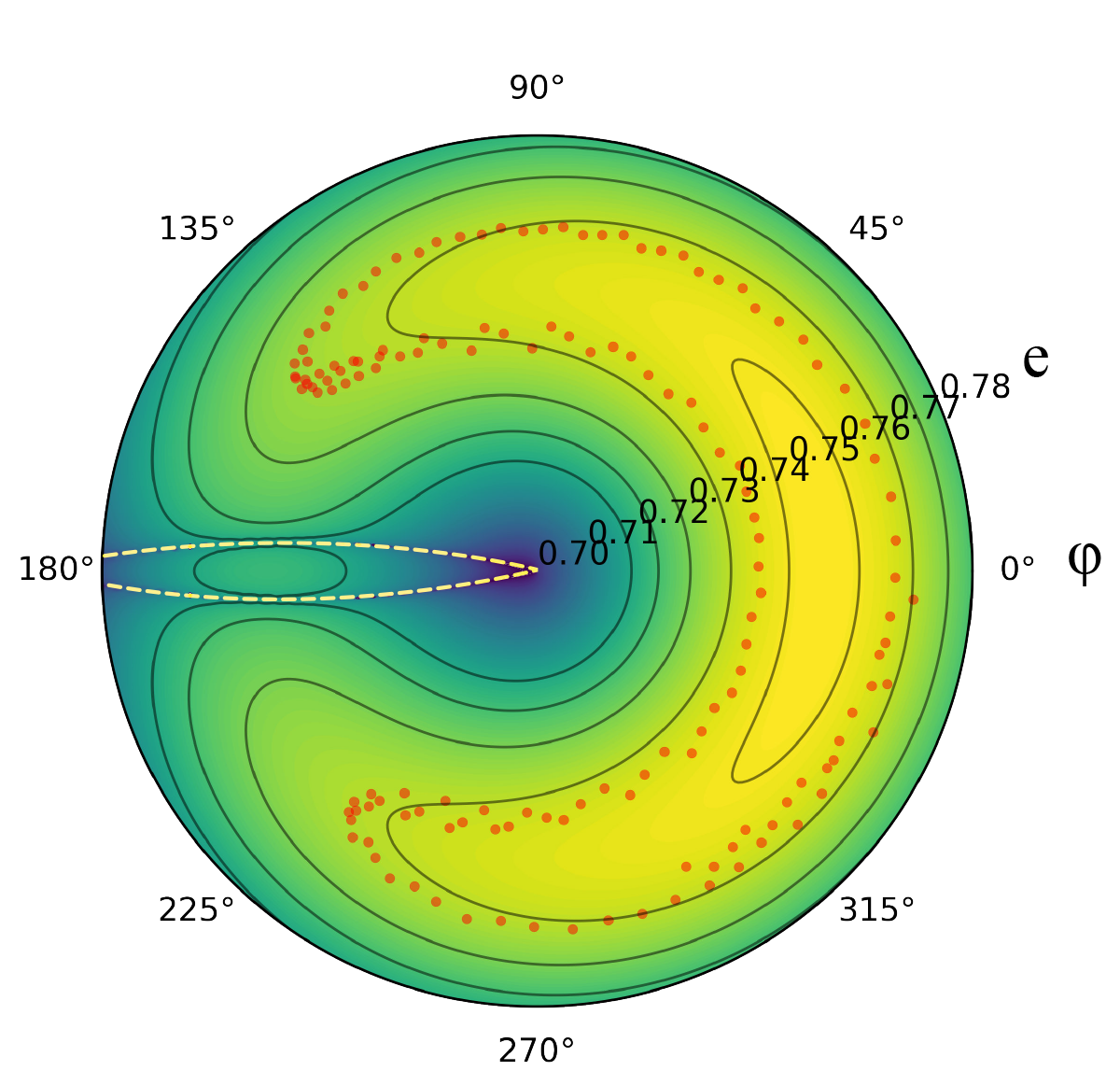}
	\caption{The phase-space portrait of the 1/-1 mean motion resonance with Saturn associated with the orbital elements of the Centaur 2006 RJ2, presented in the  $e-\phi$ polar space. The basic
	meaning of this figure is the same with Figure~\ref{fig7}. But we confine the plot range of the eccentricity to (0.7,0.78) for better visual presentation.}
\label{fig2}
\end{figure}

\begin{figure}
	\includegraphics[width=\columnwidth]{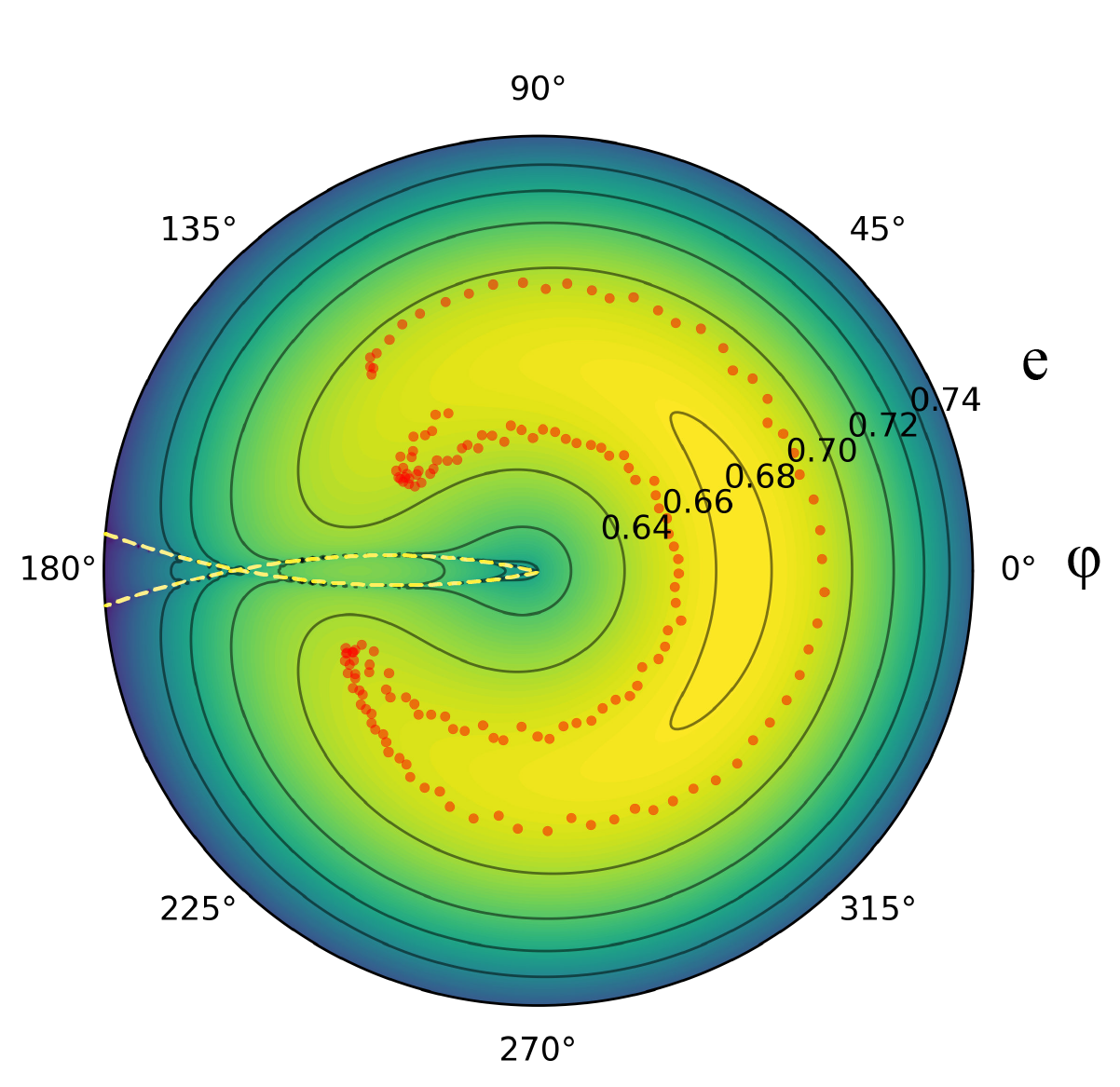}
	\caption{The phase-space portrait of the 1/-1 mean motion resonance with Saturn associated with the orbital elements of the Centaur 2006 BZ8. The basic
	meaning of this figure is the same with Figure~\ref{fig7}. But we confine the plot range of the eccentricity to (0.62,0.75) for better visual presentation.}
\label{fig6}
\end{figure}

As shown in Tables~\ref{tab:1} and ~\ref{tab:2}, the inclinations of 2006 RJ2 and 2006 BZ8 are both close to 180\degr.
We can apply this planar semi-analytical model to these two {Centaurs} to evaluate their dynamical motions. The $e-\phi$ diagram for
2006 RJ2 is presented in Figure~\ref{fig2}. We notice that the motion of 2006 RJ2 in the numerical model is predicted pretty accurately by the planar semi-analytical model. The motion of
the {Centaur} is just around the ideal equilibrium point of the 1/-1 resonance. The curves given by the numerical integration
correspond to the large-amplitude libration as defined by \citet{Murray:396121}.
The phase-space portrait of 2006 BZ8 after capturing in 1/-1 resonance also presented in Figure~\ref{fig6}. The motion in numerical model is not
predicted pretty accurately as 2006 RJ2 due to the non-negligible impact of the Jupiter. But it also belongs to the scale of large-amplitude libration around the ideal equilibrium point.
As shown in Figures~\ref{fig7}-\ref{fig6}, the collision curve slices the entire phase space into different region.
This work further verify the 1/-1 resonant state of 2006 RJ2 and 2006 BZ8.
We will research on the theory of the retrograde mean motion resonance that can apply to three-dimensional cases in future work.

\section{Conclusions}
\citet{Morais2016} predicted stable co-orbit resonances exist at all inclinations, but an asteroid in stable retrograde co-orbit resonance with planet
has not yet been discovered until the first identified asteroid in 1/-1 resonance with Jupiter, 2015 BZ509 \citep{Wiegert2017}. In this paper, we identify four potential retrograde co-orbital {Centaurs} of Saturn.
The statistical results of 1000 clones (include nominal orbit) of each candidate by numerical simulations are given in Table~\ref{tab:3}.
According to our statistical results, 2006 RJ2 is the best candidate to be currently in a 1/-1 mean motion resonance with Saturn, as another minor body found in retrograde co-orbit resonance with the planet after 2015 BZ509.
2017 SV13 is another important potential candidate whose clones are are always captured in this particular resonant state during the whole integration timespan. Moreover, 2012 YE8 and 2006 BZ8 are also Centaurs of interest but their current and long-term 1/-1 resonant state with Saturn is less likely.


Our findings suggest that {small bodies} in retrograde co-orbit resonance with giant planets may be more common in our solar system than previously expected.
The existence of these mysterious {small bodies} may give us a kind of unique perspective to study the co-orbit motion in our solar system.
{As for the small bodies on planet-crossing orbits such as Centaurs}, \citet{doi:10.1093/mnras/stw1532, BaileyBrenaeL2009, Namouni2015} suggested that those locked in stable resonances
have longer lifetimes than others.
That is to say, the {Centaurs} in retrograde co-orbit resonance we identified here, 2006 RJ2, 2006 BZ8, 2012 YE8, 2017 SV13 may belong to the most
dynamically stable objects among Centaurs.
This is significant for the research of the Centaurs' origin and dynamical evolution.

The dynamics of the planar retrograde co-orbit resonance is studied through the averaged Hamiltonian.
We introduce an useful integrable approximation for planar 1/-1 resonance and confirm the 1/-1 resonant states by the comparison of
the phase-space portrait between the semi-analytical model and the numerical results.
The planar 1/-1 resonance semi-analytical model can predict the motion of 2015 BZ509 (its inclination is close to 180\degr) quite well.
The real motions of the 2006 RJ2 and 2006 BZ8 (after capturing in 1/-1 resonance) in the solar system are just around the ideal equilibrium points of the 1/-1 resonance given by the planar semi-analytical model.
These results further verify our findings given by the numerical simulations.
\citet{Namouni2017} confirmed that the efficiency of co-orbit capture at large retrograde inclination is an intrinsic feature of co-orbital resonance.
Therefore, this planar semi-analytical model can provide a good approximation to the real motion of most small bodies in 1/-1 resonance with the planets.

Centaurs are small bodies of the solar system on the long-term unstable orbits with a semi-major axis between those of the outer planets.
As a result, these retrograde Centaurs wander radially between the outer planets have lots of "chance" to be captured in co-orbit resonance because of the
frequent close encounters with giants planets.

The conclusions of our present study encourage the search for such minor bodies on a larger scale to analyze the origins and evolution of these mysterious objects.
Our results imply that retrograde mean motion resonance plays a crucial role in the dynamical evolution of the Centaurs on retrograde orbits.
In this regard, the findings of this paper are useful for understanding the origin and dynamical evolution of the Centaurs and Damocloids on retrograde orbits.



\begin{acknowledgements}
  This work was supported by the National Natural Science
  Foundation of China (Grant No.11772167)
\end{acknowledgements}

%
  \bibliographystyle{aa} 
  \bibliography{ref} 
%











\end{document}